\documentstyle[amssymb,preprint,prd,aps]{revtex}
\global\arraycolsep=2pt
\makeatletter
\def\fmslash{\@ifnextchar[{\fmsl@sh}{\fmsl@sh[0mu]}}
\def\fmsl@sh[#1]#2{  \mathchoice
    {\@fmsl@sh\displaystyle{#1}{#2}}    {\@fmsl@sh\textstyle{#1}{#2}}    
{\@fmsl@sh\scriptstyle{#1}{#2}}    {\@fmsl@sh\scriptscriptstyle{#1}{#2}}}
\def\@fmsl@sh#1#2#3{\m@th\ooalign{$\hfil#1\mkern#2/\hfil$\crcr$#1#3$}}
\makeatother

\begin{document}
\draft
\title{Distinguishing a set of full product bases needs only projective
measurements and classical communication }
\author{Ping-Xing Chen$^{1,2}${\footnotesize \thanks{%
E-mail: pxchen@nudt.edu.cn}} and Cheng-Zu Li$^2$}
\address{1. Laboratory of Quantum Communication and Quantum Computation, \\
University of Science and Technology of\\
China, Hefei, 230026, P. R. China.\\
2. Department of Applied Physics, National University of\\
Defense Technology,\\
Changsha, 410073, \\
P. R. China. }
\date{\today}
\maketitle

\begin{abstract}
Nonlocality without entanglement is an interesting field. A manifestation of
quantum nonlocality without entanglement is the local indistinguishability
of a set of orthogonal product states. In this paper we analyze the
character of operators to distinguish a set of full product bases in a
multi-partite system, and show that distinguishing perfectly a set of full
product bases needs only local projective measurements and classical
communication, and these measurements cannot damage each product basis.
Employing these conclusions one can discuss local distinguishability of full
product bases easily. Finally we discuss the generalization of these results
to the locally distinguishability of a set of incomplete product bases.
\end{abstract}


\thispagestyle{empty}

\newpage \pagenumbering{arabic} 
%%%%%%%%%%%%%%%%%%%%%%%%%%%%%%%%%%%%%%%%%%%%%%%%%%%%%%%%%%%%%%%%%%%%%%%%%%%%%%%%

An important manifestation of quantum nonlocality is entanglement \cite{1}.
The entangled states can be used for novel forms of information processing,
such as quantum cryptography \cite{2,3}, quantum teleportation \cite{4}, and
fast quantum computation \cite{5}. However, there also exists nonlocality in
inentangled states \cite{6,7}, even in a state of a particle \cite{3}. This
is known as nonlocality without entanglement. The protocols of single photon
cryptography \cite{3} are examples which uses the nonlocality without
entanglement. The nonlocality without entanglement may be an important field
just as the entanglement. Closely related to the nonlocality without
entanglement is the local distinguishability of a set of inentangled states 
\cite{6,7}.

Alice, Bob and Charles et al share a quantum system, in one of a known set
of possible orthogonal states. They do not, however, know which state they
have. These states are locally distinguishable if there are some sequence of
local operations and classical communication (LOCC) by which Alice, Bob and
Charles et al can always determine which state they own. There are many
interesting works on the local distinguishability of orthogonal states \cite
{6,7,8,9,10,11,12,13,14,15}. These works improve our understanding on
nonlocality. The discussion on the local distinguishability of orthogonal
product states (OPSs) may enlarge our acknowledge of nonlocality without
entanglement. Bennett et al first \cite{6} showed that there are 9 OPSs in a 
$3\otimes 3$ system which are indistinguishable by LOCC. Walgate et al \cite
{7} provide a more simple proof of indistinguishability of the Bennett's 9
OPSs. However few papers discussed the local distinguishability of more
general OPSs in a multi-partite system.

This paper will focus on the local distinguishability of a set of complete
OPSs $\{\left| \Psi _k\right\rangle \}$ in a multi-partite system. We will
show that a set of full OPSs are LOCC perfectly distinguishable if and only
if these OPSs are distinguishable by projective measurements and classical
communication, and these measurements cannot damage each state $\left| \Psi
_k\right\rangle .$ Using this result we can prove easily that the Bennett's
9 OPSs \cite{6} are indistinguishable by LOCC, and can provide many new sets
of locally indistinguishable OPSs in multi-partite systems. Finally we
discuss the generalization of these results to the LOCC distinguishability
of incomplete product bases.

Alice, Bob and Charles et al share a quantum system which may be in one of
the possible states $\{\left| \psi _k\right\rangle ,k=1,\cdots ,M\}$. Any
protocol to distinguish these possible states can be conceived as successive
rounds of measurements and communication by Alice, Bob and Charles et al.
After $N$ $(N\geqslant 1)$ rounds of measurements and communication, there
are many possible outcomes which correspond to many measurement operators $%
\{A_{iN}\otimes B_{iN}\otimes C_{iN}\otimes \cdots \}$ acting on the Alice,
Bob and Charles's Hilbert space. Each of these operators is a product of the
positive operators and unitary maps corresponding to Alice's, Bob's and
Charles's measurement and rotations, and represents the effect of the N
measurements and communication. If the outcome $iN$ occurs, the given state $%
\left| \psi _k\right\rangle $ becomes \cite{7,16}:

\begin{equation}
\left| \psi _k\right\rangle \rightarrow A_{iN}\otimes B_{iN}\otimes
C_{iN}\otimes \cdots \left| \psi _k\right\rangle .  \label{f}
\end{equation}
Operator $A_{iN},B_{iN},C_{iN}$ can be expressed as (see Appendix in this
paper, or Ref.\cite{17}):

\[
A_{iN}=c_1^{iN}\left| \phi _1^{\prime iN}\right\rangle \left\langle \phi
_1^{iN}\right| +\cdots +c_{n_a^{iN}}^{iN}\left| \phi _{n_a^{iN}}^{\prime
iN}\right\rangle \left\langle \phi _{n_a^{iN}}^{iN}\right| ; 
\]
\begin{equation}
B_{iN}=d_1^{iN}\left| \xi _1^{\prime iN}\right\rangle \left\langle \xi
_1^{iN}\right| +\cdots +d_{n_b^{iN}}^{iN}\left| \xi _{n_b^{iN}}^{\prime
iN}\right\rangle \left\langle \xi _{n_b^{iN}}^{iN}\right| ;  \label{0}
\end{equation}
\[
C_{iN}=e_1^{iN}\left| \eta _1^{\prime iN}\right\rangle \left\langle \eta
_1^{iN}\right| +\cdots +e_{n_c^{iN}}^{iN}\left| \eta _{n_c^{iN}}^{\prime
iN}\right\rangle \left\langle \eta _{n_c^{iN}}^{iN}\right| ; 
\]
where $\left\{ \left| \phi _j^{\prime iN}\right\rangle ,\text{ }j=1,\cdots
,n_a^{iN}\right\} ,\left\{ \left| \phi _j^{iN}\right\rangle ,\text{ }%
j=1,\cdots ,n_a^{iN}\right\} $ are Alice's two set of orthogonal vectors; $%
\left\{ \left| \xi _l^{\prime iN}\right\rangle ,\text{ }l=1,\cdots
,n_b^{iN}\right\} ,\left\{ \left| \xi _l^{iN}\right\rangle ,\text{ }%
l=1,\cdots ,n_b^{iN}\right\} $ are Bob's two set of orthogonal vectors; $%
\left\{ \left| \eta _p^{\prime iN}\right\rangle ,\text{ }p=1,\cdots
,n_c^{iN}\right\} ,\left\{ \left| \eta _p^{iN}\right\rangle ,\text{ }%
p=1,\cdots ,n_c^{iN}\right\} $ are Charles's two set of orthogonal vectors. $%
0\leq c_j^{iN}\leq 1,$ $j=1,\cdots ,n_a^{iN};0\leq d_l^{iN}\leq 1,$ $%
l=1,\cdots ,n_b^{iN};0\leq e_p^{iN}\leq 1,p=1,\cdots ,n_c^{iN}.$ The
operator $A_{iN}$ in (\ref{0}) can be carried out by following three
operators: 1) a projective operator which projects out $\left| \phi
_j^{iN}\right\rangle ,$ $j=1,\cdots ,n_a^{iN};$ 2) a local filter operator
which changes the relative weights of the components $\left| \phi
_j^{iN}\right\rangle ,$ $j=1,\cdots ,n_a^{iN};$ 3) a local unitary operator
which transfer the Alice's vectors from $\left\{ \left| \phi
_j^{iN}\right\rangle ,\text{ }j=1,\cdots ,n_a^{iN}\right\} $ to $\left\{
\left| \phi _j^{\prime iN}\right\rangle ,\text{ }j=1,\cdots
,n_a^{iN}\right\} ,$ and similarly for $B_{iN}$ and $C_{iN}.$ Obviously, the
effect of the local unitary operator 3) is to change the bases of the
subspace projectived out by the projective operator 1), and does not affect
the distinguishability of the possible states $\{\left| \psi _k\right\rangle
,k=1,\cdots ,M\}.$ So $A_{iN}$ in (\ref{0}) can be replaced by

\begin{equation}
A_{iN}=c_1^{iN}\left| \phi _1^{iN}\right\rangle \left\langle \phi
_1^{iN}\right| +\cdots +c_{n_a^{iN}}^{iN}\left| \phi
_{n_a^{iN}}^{iN}\right\rangle \left\langle \phi _{n_a^{iN}}^{iN}\right| ,
\label{5}
\end{equation}
and similarly for $B_{iN}$ and $C_{iN}.$

Definition 1: For each operator $A_{iN}\otimes B_{iN}\otimes C_{iN}\otimes
\cdots $ in (\ref{f}), if states $\left\{ A_{iN}\otimes B_{iN}\otimes
C_{iN}\otimes \cdots \left| \psi _k\right\rangle ,k=1,\cdots ,M\right\} $
are LOCC distinguishable, we say that operator $A_{iN}\otimes B_{iN}\otimes
C_{iN}\otimes \cdots ${\it \ is effective to distinguish the states }$%
\left\{ \left| \psi _k\right\rangle \right\} .$

Theorem 1 $\left\{ \left| \Psi _k\right\rangle ,k=1,\cdots ,M\right\} $ is a
set of complete orthogonal product states in a multi-partite system. The
states $\left\{ \Psi _k\right\} $ are LOCC perfectly distinguishable if and
only if the states are distinguishable by projective measurements and
classical communication, and these measurements cannot damage each state $%
\left| \Psi _k\right\rangle .$

Proof: We first prove theorem 1 for the cases of bi-partite systems. The
sufficiency is obvious. We need only prove the necessity. Suppose that Alice
and Bob share a $n\otimes m$ system which has $nm$ possible OPSs $\{\left|
\Psi _k\right\rangle =\left| \upsilon _k\right\rangle _A\left|
y_k\right\rangle _B,k=1,\cdots ,nm\},$ where $\left| \upsilon
_k\right\rangle _A,\left| y_k\right\rangle _B$ is a state of Alice's and
Bob's, respectively. If the set of states $\left\{ \Psi _k\right\} $ is
perfectly distinguishable by LOCC, there must be a complete set of final
operators $\left\{ A_{if}\otimes B_{if}\right\} $ representing the effect of
all measurements and communication, such that if every outcome $if$ occurs
Alice and Bob know with certainty that they were given the state $\left|
\Psi _i\right\rangle \in \{\left| \Psi _k\right\rangle \}$. This means that:

\begin{eqnarray}
A_{if}\otimes B_{if}\left| \Psi _i\right\rangle &\neq &0;  \label{1} \\
A_{if}\otimes B_{if}\left| \Psi _j\right\rangle &=&0,\text{ }j\neq i. 
\nonumber
\end{eqnarray}
Operators $A_{if},B_{if}$ have similar forms as $A_{iN},B_{iN}$ in (\ref{5})
but for $N\rightarrow f.$ We note that $A_{if}\otimes B_{if}$ can
``indicates'' $\left| \Psi _i\right\rangle $ and only $\left| \Psi
_i\right\rangle $ \cite{15}$.$ Since the number of the operators satisfying
the (\ref{1}) is bigger than 1, a state $\left| \Psi _i\right\rangle $ can
be ``indicated'' by more than a operator, in general. By the general
expressions $A_{if},B_{if}$ as in (\ref{5}) if operator $A_{if}\otimes
B_{if} $ can ``indicates'' $\left| \Psi _i\right\rangle $ and only $\left|
\Psi _i\right\rangle ,$ i.e., (\ref{1}) holds, the state $\left| \Psi
_i\right\rangle $ should contain all or part of orthogonal product vectors
in the following:

\begin{equation}
\left| \phi _1^{if}\right\rangle \left| \xi _1^{if}\right\rangle ,\cdots
,\left| \phi _1^{if}\right\rangle \left| \xi _{n_b^{if}}^{if}\right\rangle
,\cdots ,\left| \phi _{n_a^{if}}^{if}\right\rangle \left| \xi
_1^{if}\right\rangle ,\cdots ,\left| \phi _{n_a^{if}}^{if}\right\rangle
\left| \xi _{n_b^{if}}^{if}\right\rangle ,  \label{44}
\end{equation}
and $\left| \Psi _j\right\rangle $ $(j\neq i)$ do not contain any product
vectors in (\ref{44}), i.e., each product vectors in (\ref{44}) is
orthogonal to $\left| \Psi _j\right\rangle $ $(j\neq i).$ Since for a $%
n\otimes m$ system the vector which is orthogonal to $nm-1$ orthogonal
states $\left| \Psi _j\right\rangle $ $(j\neq i)$ is alone, if operator $%
A_{if}\otimes B_{if}$ ``indicates'' $\left| \Psi _i\right\rangle ,$ then $%
\left| \Psi _i\right\rangle $ contains only one product vector in (\ref{44}%
). Namely, $\left| \Psi _i\right\rangle $ should be one of the product
vectors in (\ref{44}). So all operators $A_{if}\otimes B_{if}$
``indicating'' $\left| \Psi _i\right\rangle $ and only $\left| \Psi
_i\right\rangle $ should project out a intact state $\left| \Psi
_i\right\rangle ,$ but not a component of $\left| \Psi _i\right\rangle ,$
and then all OPS in the set $\{\left| \Psi _k\right\rangle \}$ are
eigenvectors of each operator in operators $\{A_{if}\otimes B_{if}\}.$ For
the same reason, we can prove that all OPSs in the set $\{\left| \Psi
_k\right\rangle \}$ are eigenvectors of each operator $A_{iN}\otimes B_{iN}$
representing the effect of N round measurements and communication, $%
N=1,2,\cdots .$ In fact, suppose that during the first measure to
distinguish the states $\{\left| \Psi _k\right\rangle \}$ $($suppose Alice
do the first measure) if an {\it effective }operator $A_{i1}\ $does not
project out an intact OPS, but a part of a OPS, then the two orthogonal
parts of the OPS are orthogonal to $nm-1$ orthogonal states. This is
impossible. On the other hand, after Alice and Bob finished the first
measure and get a outcome, the whole space collapses into a subspace and the
OPSs in this subspace form a set of complete bases of the subspace. So the
sequent measures have same property as the first measure.

Let's now prove that the set of final operators $\{A_{if}\otimes B_{if}\}$
can be carried out by projective measurements and classical communication.
To achieve this, we consider the Alice's first measurement described by $%
\{A_{i1}\}$

\begin{equation}
A_{i1}=c_1^{i1}\left| \phi _1^{i1}\right\rangle _A\left\langle \phi
_1^{i1}\right| +\cdots +c_{n_a^{i1}}^{i1}\left| \phi
_{n_a^{i1}}^{i1}\right\rangle _A\left\langle \phi _{n_a^{i1}}^{i1}\right| .
\label{13}
\end{equation}
Operator $A_{i1}$ project out a subspace of Alice spanned by Alice's bases $%
\left| \phi _1^{i1}\right\rangle _A,\cdots ,\left| \phi
_{n_a^{i1}}^{i1}\right\rangle _A$. This subspace should contain some intact
Alice's vectors of the OPSs (since $A_{i1}$ should project out some intact
OPSs). Since all OPSs $\{\left| \Psi _k\right\rangle \}$ are eigenvectors of
the operator $A_{i1},$ if $\{A_{i1}\}$ is {\it effective to distinguish the
states }$\{\left| \Psi _k\right\rangle \}$ then so does operators $%
\{A_{i1}^{\prime }\}:$

\begin{equation}
A_{i1}^{\prime }=\left| \phi _1^{i1}\right\rangle \left\langle \phi
_1^{i1}\right| +\cdots +\left| \phi _{n_a^{i1}}^{i1}\right\rangle
\left\langle \phi _{n_a^{i1}}^{i1}\right| .  \label{6}
\end{equation}
Operator $A_{i1}^{\prime }$ projects out a same subspace $\varkappa _i$ as $%
A_{i1}$ does, and all OPSs $\{\left| \Psi _i\right\rangle \}$ are the
eigenvectors of the operator $A_{i1}^{\prime }.$ If operators $%
\{A_{i1}^{\prime }\}$ is not a set of projective operators, we can find a
set of projective operators $\{A_{i1}^{\prime \prime }\}$ by following
protocol such that if $\{A_{i1}^{\prime }\}$ is {\it effective to
distinguish the states }$\{\left| \Psi _k\right\rangle \}$ then so does
operators $\{A_{i1}^{\prime \prime }\}$. We first choose two operators $%
A_{11}^{\prime },A_{21}^{\prime }$ described by $A_{i1}^{\prime }(i=1,2)$ in
(\ref{6}). Operators $A_{11}^{\prime },A_{21}^{\prime }$ projects out
subspace $\varkappa _1$, $\varkappa _2$, respectively. Both $\varkappa _1$
and $\varkappa _2$ should contains intact Alice's vectors of some OPSs $%
\left| \Psi _k\right\rangle .$ Suppose $\varkappa _1$ and $\varkappa _2$
contains the Alice's vectors of $\left| \Psi _{k1}\right\rangle s$ and $%
\left| \Psi _{k2}\right\rangle s,$ respectively, $\left| \Psi
_{k1}\right\rangle s$, $\left| \Psi _{k2}\right\rangle s\in \{\left| \Psi
_k\right\rangle \}.$ The Alice's vectors of the OPSs belonging to $\left|
\Psi _{k2}\right\rangle s$ but not to $\left| \Psi _{k1}\right\rangle s$
form a subspace $\varkappa $ of $\varkappa _2.$ Obviously, $\varkappa $ is
orthogonal to $\varkappa _1.$ We note operator $A_{21}^{\prime \prime }$
projects out subspace $\varkappa $ and only $\varkappa $ ( In fact, $%
\varkappa _2$ $=\varkappa $ $\cup (\varkappa _2\cap \varkappa _1),$ and $%
\varkappa $ is orthogonal to $(\varkappa _2\cap \varkappa _1).$ Operator $%
A_{21}^{\prime }$ projects out subspace $\varkappa _2.$ The effect of
operator $A_{21}^{\prime \prime }$ is to discard some bases of $\varkappa
_2, $ and to project out subspace $\varkappa $). If we replace $%
A_{21}^{\prime } $ by $A_{21}^{\prime \prime },$ then $A_{11}^{\prime }$ and 
$A_{21}^{\prime \prime }$ project out some OPSs as same as $A_{11}^{\prime }$
and $A_{21}^{\prime }$ do, and these OPSs are distinguishable by LOCC if $%
A_{11}^{\prime }$ and $A_{21}^{\prime }$ are {\it effective to distinguish
the states }$\{\left| \Psi _k\right\rangle \}$. Similarly, operator $%
A_{31}^{\prime }$ projects out subspace $\varkappa _3.$ we can discard the
bases of $\varkappa _3$ which are vectors of subspace $\varkappa _1$ or $%
\varkappa _2.$ The left bases of $\varkappa _3$ span a subspace projected
out by operator $A_{31}^{\prime \prime }.$ By a sequence of similar
operations we find always a set of orthogonal projective operators $%
\{A_{11}^{\prime },A_{21}^{\prime \prime },A_{31}^{\prime \prime },\cdots \}$
such that if operators $\{A_{i1}^{\prime }\}$ are {\it effective to
distinguish the OPSs}, so do operators $\{A_{11}^{\prime },A_{21}^{\prime
\prime },A_{31}^{\prime \prime },\cdots \}$. Obviously $A_{11}^{\prime
+}A_{11}^{\prime }+A_{21}^{\prime \prime +}A_{21}^{\prime \prime
}+A_{31}^{\prime \prime +}A_{31}^{\prime \prime }+\cdots =I.$ So Alice's
first measure can be carried out by a set projective operators $%
\{A_{11}^{\prime },A_{21}^{\prime \prime },A_{31}^{\prime \prime },\cdots \}$%
, similarly for Bob's first measure. On the other hand, after Alice and Bob
finished the first measure and get a outcome, the whole space collapses into
a subspace and the OPSs in this subspace form a set of complete bases of the
subspace. So the sequent measures have same property as the first measure.
Thus the set of operators $\{A_{if}\otimes B_{if}\}$ can be carried out by
projective measurements and classical communication. The whole proof is
completely fit to the cases of multi-partite systems. This ends the proof.

Definition 2: If two states $\left| \Phi _1\right\rangle $ and $\left| \Phi
_2\right\rangle $ satisfying $\langle \Phi _1\left| \Phi _2\right\rangle
\neq 0,$ we say $\left| \Phi _1\right\rangle ${\it \ and }$\left| \Phi
_1\right\rangle ${\it \ are relative}. This is noted as $\left| \Phi
_1\right\rangle \longleftrightarrow \left| \Phi _2\right\rangle ;$ if $%
\left| \Phi _1\right\rangle \longleftrightarrow \left| \Phi _2\right\rangle
\longleftrightarrow \cdots \longleftrightarrow \left| \Phi _N\right\rangle
(i.e.,\langle \Phi _1\left| \Phi _2\right\rangle \neq 0;\langle \Phi
_2\left| \Phi _3\right\rangle \neq 0;\cdots ;\langle \Phi _{N-1}\left| \Phi
_N\right\rangle \neq 0),$ we say $\left| \Phi _1\right\rangle ,\left| \Phi
_2\right\rangle ,\cdots ,\left| \Phi _N\right\rangle ${\it \ are relative}.

Theorem 2: A set of states $\{\left| \upsilon _i\right\rangle _A\left|
y_i\right\rangle _B,i=1,\cdots ,nm\}$ is $nm$ OPSs in a $n\otimes m$ system.
If for every given state $\left| \upsilon _i\right\rangle _A\left|
y_i\right\rangle _B,$ there are $n-1$ states $\left| \upsilon _j^{\prime
}\right\rangle _A\left| y_j^{\prime }\right\rangle _B\in \{\left| \upsilon
_i\right\rangle _A\left| y_i\right\rangle _B,i=1,\cdots ,nm\},j=1,\cdots
,n-1 $ such that

\begin{equation}
\left| \upsilon _i\right\rangle _A\longleftrightarrow \left| \upsilon
_1^{\prime }\right\rangle _A\longleftrightarrow \cdots \longleftrightarrow
\left| \upsilon _{n-1}^{\prime }\right\rangle _A,  \label{8}
\end{equation}
and $\left| \upsilon _i\right\rangle _A,\left| \upsilon _1^{\prime
}\right\rangle _A,\cdots ,\left| \upsilon _{n-1}^{\prime }\right\rangle $
are linearly independent; If for every given state $\left| \upsilon
_i\right\rangle _A\left| y_i\right\rangle _B,$ there are $m-1$ states $%
\left| \upsilon _k^{\prime }\right\rangle _A\left| y_k^{\prime
}\right\rangle _B\in \{\left| \upsilon _i\right\rangle _A\left|
y_i\right\rangle _B,i=1,\cdots ,nm\},k=1,\cdots ,m-1$ such that

\begin{equation}
\left| y_i\right\rangle _B\longleftrightarrow \left| y_1^{\prime
}\right\rangle _B\longleftrightarrow \cdots \longleftrightarrow \left|
y_{m-1}^{\prime }\right\rangle _B,  \label{9}
\end{equation}
and $\left| y_i\right\rangle _B,\left| y_1^{\prime }\right\rangle _B,\cdots
,\left| y_{m-1}^{\prime }\right\rangle _B$ are linearly independent, then
states $\{\left| \upsilon _i\right\rangle _A\left| y_i\right\rangle
_B,i=1,\cdots ,nm\}$ are not LOCC distinguishable.

Proof: Suppose Alice do the first measure (Alice goes first \cite{7}). From
theorem 1 it follows that all states $\{\left| \upsilon _i\right\rangle
_A\left| y_i\right\rangle _B,i=1,\cdots ,nm\}$ are eigenstates of Alice's
first measure described as $A_j$ in (\ref{6}). If $\left| \upsilon
_i\right\rangle _A$ is a eigenstate of operator $A_j$ with non-zero
eigenvalue, equation (\ref{8}) implies that $\left| \upsilon _1^{\prime
}\right\rangle _A$ should be also a eigenstate of operator $A_j$ with
non-zero eigenvalue, and so does $\left| \upsilon _j^{\prime }\right\rangle
_A$, $j=2,\cdots ,n-1.$ So the rank of operator $A_j$ is full. A
full-rank-operator $A_j$ would project out all OPSs and can do nothing to
distinguish states $\{\left| \upsilon _i\right\rangle _A\left|
y_i\right\rangle _B,i=1,\cdots ,nm\},$ and similarly for Bob's first
measure. So the states $\{\left| \upsilon _i\right\rangle _A\left|
y_i\right\rangle _B,i=1,\cdots ,nm\}$ are not LOCC distinguishable. This
ends the proof.

Theorem 2 above can be generalized into multi-partite cases, obviously. From
the theorem 2 we can get many cases of indistinguishable states. There are
three examples in the following:

Case 1 The 9 OPSs in a $3\otimes 3$ system in the following are
indistinguishable as shown in paper \cite{6} of Bennett et al.

\begin{eqnarray}
\left| \Psi _1\right\rangle &=&\left| 1\right\rangle _A\left| 1\right\rangle
_B;\left| \Psi _{2,3}\right\rangle =\left| 3\right\rangle _A\left| 3\pm
1\right\rangle _B;  \label{10} \\
\left| \Psi _{4,5}\right\rangle &=&\left| 2\right\rangle _A\left| 1\pm
2\right\rangle _B;\left| \Psi _{6,7}\right\rangle =\left| 3\pm
1\right\rangle _A\left| 2\right\rangle _B;  \nonumber \\
\left| \Psi _{8,9}\right\rangle &=&\left| 1\pm 2\right\rangle _A\left|
3\right\rangle _B  \nonumber
\end{eqnarray}

Case 2 The following 16 OPSs in a $4\otimes 4$ system are indistinguishable.

\begin{eqnarray}
\left| \Psi _{1,2}\right\rangle &=&\left| 1\right\rangle _A\left| 1\pm
2\right\rangle _B;\qquad \left| \Psi _{3,4}\right\rangle =\left|
2\right\rangle _A\left| 2\pm 3\right\rangle _B;  \label{11} \\
\left| \Psi _{5,6}\right\rangle &=&\left| 3\right\rangle _A\left| 3\pm
4\right\rangle _B;\qquad \left| \Psi _{7,8}\right\rangle =\left|
4\right\rangle \left| _A1\pm 4\right\rangle _B;  \nonumber \\
\left| \Psi _{9,10}\right\rangle &=&\left| 1\pm 2\right\rangle _A\left|
4\right\rangle _B;\qquad \left| \Psi _{11,12}\right\rangle =\left| 3\pm
4\right\rangle _A\left| 2\right\rangle _B;  \nonumber \\
\left| \Psi _{13,14}\right\rangle &=&\left| 2\pm 3\right\rangle _A\left|
1\right\rangle _B;\qquad \left| \Psi _{15,16}\right\rangle =\left| 1\pm
4\right\rangle _A\left| 3\right\rangle _B.  \nonumber
\end{eqnarray}

Case 3 The following 64 OPSs in a $4\otimes 4\otimes 4$ system are
indistinguishable.

\begin{eqnarray}
\left| \Psi _i\right\rangle &=&\left| \Psi _i\right\rangle _{AB}\left|
1\right\rangle _C;\qquad \left| \Psi _{i+16}\right\rangle =\left| \Psi
_i\right\rangle _{AB}\left| 2\right\rangle _C;  \label{12} \\
\left| \Psi _{i+32}\right\rangle &=&\left| \Psi _i\right\rangle _{AB}\left|
3\right\rangle _C;\qquad \left| \Psi _{i+48}\right\rangle =\left| \Psi
_i\right\rangle _{AB}\left| 4\right\rangle _C,  \nonumber \\
i &=&1,\cdots ,16  \nonumber
\end{eqnarray}
Where $\left| \Psi _i\right\rangle $ is a state in case 2.

Employing the above theorem 2, we can prove the cases 1 and 2 easily. In
case 3, Charles can do the first projective measurement by operators $\left|
1\right\rangle \left\langle 1\right| ,\left| 2\right\rangle \left\langle
2\right| ,\left| 3\right\rangle \left\langle 3\right| ,\left| 4\right\rangle
\left\langle 4\right| .$ But after Charles's first round measurement the
states of Alice and Bob' part will collapse into some indistinguishable OPSs 
$\left| \Psi _i\right\rangle _{AB}s$. So the OPSs in case 3 is
indistinguishable by LOCC.

Locally distinguishing a set of full OPSs needs only local projective
measurements and classical communication. Can this conclusion be generalized
into a set of incomplete OPSs? A set of following OPSs shows it is not
always true. Nine OPSs \cite{18}

\begin{eqnarray}
\left| \Psi _{1,2,3}\right\rangle &=&\left| \Psi _{1,2,3}\right\rangle
_{AB}\left| x\right\rangle _C;  \label{14} \\
\left| \Psi _{4,5,6}\right\rangle &=&\left| \Psi _{4,5,6}\right\rangle
_{AB}\left| y\right\rangle _C;  \nonumber \\
\left| \Psi _{7,8,9}\right\rangle &=&\left| \Psi _{7,8,9}\right\rangle
_{AB}\left| z\right\rangle _C,  \nonumber
\end{eqnarray}
where $\left| \Psi _i\right\rangle (i=1,\cdots ,9)$ is a state in (\ref{10}%
), $\left| x\right\rangle =\left| 1\right\rangle ;\left| y\right\rangle
=(\left| 1\right\rangle +\sqrt{3}\left| 2\right\rangle )/2;\left|
z\right\rangle =(\left| 1\right\rangle -\sqrt{3}\left| 2\right\rangle )/2,$
can be distinguished by Charles doing the first measure described by
operators $C_{i1}(i=1,2,3)$ \cite{18}

\begin{eqnarray}
C_{11} &=&\sqrt{\frac 23}|x^{*}><x^{*}|;  \label{15} \\
C_{21} &=&\sqrt{\frac 23}|y^{*}><y^{*}|;  \nonumber \\
C_{31} &=&\sqrt{\frac 23}|z^{*}><z^{*}|,  \nonumber
\end{eqnarray}
where $\langle $ $x|x^{*}>=\langle $ $y|y^{*}>=\langle $ $z|z^{*}>=0,$ and $%
\sum_{i=1}^3C_{i1}^{+}C_{i1}=1.$ After Charles get a outcome, 9 OPSs in (\ref
{14}) collapse into locally distinguishable 6 OPSs. However, states $\left|
\Psi _i\right\rangle (i=1,\cdots ,9)$ in (\ref{14}) cannot be distinguished
by local projective measurements and classical communication.

In conclusion, we analyze the character of operators to distinguish a set of
full OPSs in a multi-partite system, and show that to distinguish perfectly
a set of full bases needs only local projective measurements and classical
communication, and these measurements cannot damage each OPS. Employing
these conclusions one can discuss local distinguishability of full product
bases easily. An open question is that whether these conclusions can be
generalized to the local distinguishability of states in a quantum system,
the sum of Schmidt number of the states is equal to the dimensions of
Hilbert space of the system. Another open question is that which classes of
operators can be carried out only local projective measurements, since local
projective measurements are easier to be achieved than generalized POV
measurements.

\acknowledgments    We would like to thank J. Finkelstein for presenting us
a set of special states in (\ref{14}) and Guangcan Guo for his help to this
work.

Appendix

Now we will prove that operator $A_{iN},B_{iN},C_{iN}$ can be expressed as
the form in (\ref{0}). To this end, note that we can always write an
arbitrary operator $A_{iN}$ in the form \cite{17}

\begin{equation}
A_{iN}=\left| \beta _1^{iN}\right\rangle \left\langle \alpha _1^{iN}\right|
+\cdots +\left| \beta _{n_a^{iN}}^{iN}\right\rangle \left\langle \alpha
_{n_a^{iN}}^{iN}\right| ,  \label{A1}
\end{equation}
where $\left\{ \left| \alpha _1^{iN}\right\rangle ,\cdots ,\left| \alpha
_{n_a^{iN}}^{iN}\right\rangle \right\} $ is a set of Alice's orthogonal and
normalized vectors; $\left\{ \left| \beta _1^{iN}\right\rangle ,\cdots
,\left| \beta _{n_a^{iN}}^{iN}\right\rangle \right\} $ is a set of linearly
independent (possibly unnormalized) Alice's vectors \cite{19}. We take
another set of orthogonal and normalized vectors $\left\{ \left| \alpha
_1^{\prime iN}\right\rangle ,\cdots ,\left| \alpha _{n_a^{iN}}^{\prime
iN}\right\rangle \right\} $ satisfying 
\begin{equation}
\left[ \left| \alpha ^{iN}\right\rangle \right] =\left[ u\right] \left[
\left| \alpha ^{\prime iN}\right\rangle \right] ,  \label{A2}
\end{equation}
where 
\begin{equation}
\left[ \left| \alpha ^{iN}\right\rangle \right] =\left[ 
\begin{array}{c}
\left| \alpha _1^{iN}\right\rangle \\ 
\vdots \\ 
\left| \alpha _{n_a^{iN}}^{iN}\right\rangle
\end{array}
\right] ,  \label{A3}
\end{equation}
and similarly for $\left[ \left| \alpha ^{\prime iN}\right\rangle \right]
,\left[ \left| \beta ^{iN}\right\rangle \right] $ and $\left[ \left| \beta
^{\prime iN}\right\rangle \right] $ below$;$ $\left[ u\right] $ is a unitary
matrix. Eq. (\ref{A2}) can be expressed as explicitly 
\begin{equation}
\left| \alpha _j^{iN}\right\rangle =\sum_{k=1}^{n_a^{iN}}u_{jk}\left| \alpha
_k^{iN}\right\rangle ,\text{ }j=1,...,n_a^{iN},  \label{A33}
\end{equation}
where $u_{jk}$ is an element of the unitary matrix $\left[ u\right] $.

Taking (\ref{A2}) we can rewrite (\ref{A1}) as 
\begin{equation}
A_{iN}=\left| \beta _1^{\prime iN}\right\rangle \left\langle \alpha
_1^{\prime iN}\right| +\cdots +\left| \beta _{n_a^{iN}}^{\prime
iN}\right\rangle \left\langle \alpha _{n_a^{iN}}^{\prime iN}\right| .
\label{A4}
\end{equation}
Obviously, 
\begin{equation}
\left[ \left| \beta ^{iN}\right\rangle \right] =\left[ u\right] \left[
\left| \beta ^{\prime iN}\right\rangle \right] .  \label{A5}
\end{equation}
Eq. (\ref{A5}) can be expressed explicitly as similar as Eq.(\ref{A33}). $%
\left\{ \left| \beta _1^{iN}\right\rangle ,\cdots ,\left| \beta
_{n_a^{iN}}^{iN}\right\rangle \right\} $ is a set of physical states, the
mixture of these states, 
\begin{equation}
\rho =\sum_{j=1}^{n_a^{iN}}\left| \beta _j^{iN}\right\rangle \left\langle
\beta _j^{iN}\right|   \label{A}
\end{equation}
is a density matrix. (if $Tr\rho \neq 1,\rho $ multiplied by a constant
became a density matrix). The density matrix $\rho $ has always a set of
eigenstates $\left\{ \left| \beta _1^{^{\prime \prime }iN}\right\rangle
,\cdots ,\left| \beta _{n_a^{iN}}^{^{\prime \prime }iN}\right\rangle
\right\} ($unnormalized) such that 
\begin{equation}
\rho =\sum_{j=1}^{n_a^{iN}}\left| \beta _j^{^{\prime \prime
}iN}\right\rangle \left\langle \beta _j^{^{\prime \prime }iN}\right| .
\label{A7}
\end{equation}
According to Wootters' criterion \cite{20}, the two sets of pure states, $%
\left\{ \left| \beta _1^{^{\prime \prime }iN}\right\rangle ,\cdots ,\left|
\beta _{n_a^{iN}}^{^{\prime \prime }iN}\right\rangle \right\} $ and $\left\{
\left| \beta _1^{iN}\right\rangle ,\cdots ,\left| \beta
_{n_a^{iN}}^{iN}\right\rangle \right\} $ fulfill 
\begin{equation}
\left[ \left| \beta ^{^{\prime \prime }iN}\right\rangle \right] =\left[
A\right] \left[ \left| \beta ^{iN}\right\rangle \right] ,  \label{A8}
\end{equation}
where $\left[ A\right] $ is a matrix the columns of which form a set of
orthogonal bases. If the two sets have the same number of pure states, $%
\left[ A\right] $ is a unitary matrix (since $\left\{ \left| \beta
_1^{iN}\right\rangle ,\cdots ,\left| \beta _{n_a^{iN}}^{iN}\right\rangle
\right\} $ is a set of linearly independent vectors, $\left[ A\right] $ in (%
\ref{A8}) is a unitary matrix). So we can always find a unitary matrix so
that $\left\{ \left| \beta _1^{\prime iN}\right\rangle ,\cdots ,\left| \beta
_{n_a^{iN}}^{\prime iN}\right\rangle \right\} $ in (\ref{A5}) is a set of
orthogonal (possibly unnormalized) states. Namely, we can always express $%
A_{iN}$ as in (\ref{A4}) such that $\left\{ \left| \beta _1^{\prime
iN}\right\rangle ,\cdots ,\left| \beta _{n_a^{iN}}^{\prime iN}\right\rangle
\right\} $ and $\left\{ \left| \alpha _1^{\prime iN}\right\rangle ,\cdots
,\left| \alpha _{n_a^{iN}}^{\prime iN}\right\rangle \right\} $ is orthogonal
vectors, respectively. Normalizing the states $\left\{ \left| \beta
_1^{\prime iN}\right\rangle ,\cdots ,\left| \beta _{n_a^{iN}}^{\prime
iN}\right\rangle \right\} $ in (\ref{A4}), we get an expression of $A_{iN}$
as in (\ref{0}). Similarly for $B_{iN}$ and $C_{iN}.$

\end{document}